\documentclass[12pt, draftclsnofoot,peerreview,onecolumn]{IEEEtran}

\IEEEoverridecommandlockouts

\usepackage{color}
\usepackage{graphicx}
\usepackage{amsmath}
\usepackage{amssymb}
\usepackage{algorithm}
\usepackage{algorithmic}
\usepackage{fancyhdr}

\newcommand{\be}{\begin{equation}}
\newcommand{\ee}{\end{equation}}
\newcommand{\bea}{\begin{eqnarray}}
\newcommand{\eea}{\end{eqnarray}}
\newcommand{\ba}{\begin{array}}
\newcommand{\ea}{\end{array}}

\title{A Hardware-Efficient Hybrid Beamforming Solution for mmWave MIMO Systems }

\author{Ming Li,~\IEEEmembership{Senior Member,~IEEE,}
       Zihuang Wang,~\IEEEmembership{Student Member,~IEEE,}
       Hongyu Li,
       Qian Liu,~\IEEEmembership{Member,~IEEE,}
       and Liang Zhou

\thanks{Ming Li, Zihuang Wang, and Hongyu Li are with the School of Information and Communication Engineering, Dalian University of Technology, Dalian, Liaoning 116024, China, (e-mail: mli@dlut.edu.cn, wangzihuan@mail.dlut.edu.cn, hongyuli@mail.dlut.edu.cn).}
\thanks{Qian Liu is with the School of Computer Science and Technology, Dalian University of Technology, Dalian, Liaoning 116024, China (e-mail: qianliu@dlut.edu.cn).}
\thanks{Liang Zhou is with the Key Laboratory of Broadband Wireless Communications and Sensor Network Technology, Ministry of Education, Nanjing University of Posts and Telecommunications, Nanjing 210003, China (e-mail: liang.zhou@njupt.edu.cn).}
\thanks{This paper is supported in part by the National Natural Science Foundation of China (Grant No. 61671101, 61601080, and 61761136019) and in part by the Fundamental Research Funds for the Central Universities (Grant No. DUT17JC10 and DUT18JC17).}}

\begin{document}

\renewcommand{\baselinestretch}{1.3}

\maketitle

\thispagestyle{fancy}
\lhead{This paper is accepted to be published in IEEE Wireless Communications.}
\renewcommand{\headrulewidth}{0pt}
\setcounter{page}{1}

\pagestyle{plain}

\begin{abstract}
In millimeter wave (mmWave) communication systems, existing hybrid beamforming solutions generally require a large number of high-resolution phase shifters (PSs) to realize analog beamformers, which still suffer from high hardware complexity and power consumption. Targeting at this problem, this article introduces a novel hardware-efficient hybrid
precoding/combining architecture, which only employs a limited number of simple phase over-samplers (POSs) and a switch (SW) network to achieve maximum hardware efficiency while maintaining satisfactory spectral efficiency performance. The POS can be realized by a simple circuit and simultaneously outputs several parallel signals with different phases. With the aid of a simple switch network, the analog precoder/combiner is implemented by feeding the signals with appropriate phases to antenna arrays or RF chains.
We analyze the design challenges of this POS-SW-based hybrid beamforming architecture and present potential solutions to the fundamental issues, especially the precoder/combiner design and the channel estimation strategy.
Simulation results demonstrate that this hardware-efficient structure can achieve comparable spectral efficiency but much higher energy efficiency than that of the traditional structures.
\end{abstract}


\newpage

\renewcommand{\baselinestretch}{2.0}

\section{Introduction}
The past decade has witnessed the rapid proliferation of wireless devices along with the exponential growth of data traffic, which significantly stimulates the research and development of 5G networks. Three key technologies are suggested to 5G networks for extensive capacity enhancement: Exploiting millimeter wave (mmWave) frequency, adopting massive multiple-input multiple-output (MIMO) systems, and network densification by deploying pico-cells and femto-cells. MmWave wireless communications, operating in the frequency bands of 30-300 GHz, act as a bridge connecting these symbiotic technologies. To be more specific, the smaller wavelength of mmWave signals allows a large antenna array to be packed in a small physical dimension. This enables the deployment of massive MIMO systems for mobile devices, which is virtually impossible in the current cellular networks (e.g. large antenna arrays are typically employed at the base-stations due to the huge size). In addition, the high operating frequency of mmWave communications is extremely suitable for pico-cells and femto-cells for supporting high data rate with intended short coverage range. The narrow beampattern of mmWave MIMO systems is also very useful for mitigating the interference in the future ultra-dense networks.
Therefore, mmWave MIMO communications have been considered as a promising candidate for 5G networks to fundamentally solve the spectrum congestion problem and support high data rates \cite{Pi CM 11}-\cite{Swindlehurst 14}.

However, mmWave communications still need to overcome several technical difficulties before the real-world deployment. First of all, as a negative result of the ten-fold increase of the carrier frequency, the propagation loss in mmWave bands is much higher than that of conventional frequency bands (e.g. 2.4 GHz) due to atmospheric absorption, rain attenuation, and low penetration. Therefore, mmWave communications require tight integration of massive MIMO techniques, especially the pre/post-coding mechanisms, which can provide sufficient beamforming gain to overcome the severe propagation loss of mmWave channels.
Yet, another major difficulty of mmWave communications is the hardware constraints.
Due to much higher carrier frequency and wider bandwidth in mmWave communications, the radio frequency (RF) components with high-precision analog-to-digital converters (ADCs) and digital-to-analog converters (DACs) lead to enormous economic cost and high power consumption. Therefore, it becomes difficult to dedicate an individual RF chain for each antenna and makes the conventional full-digital beamforming infeasible in practical mmWave MIMO systems.

Recently, economical and energy-efficient analog/digital hybrid beamforming provides a promising solution by allowing a small number of RF chains.
The hybrid beamforming approaches generally adopt a large-scale high-resolution phase shifter (PS) network to implement high-dimensional analog beamformer to compensate for the severe path-loss at mmWave bands, and a few RF chains to realize low-dimensional digital beamforming to provide the necessary flexibility to perform advanced multiplexing/multiuser techniques \cite{Heath 16}.
Since the power consumption and hardware complexity of the analog beamformer are proportional to the number and the resolution of PSs, the adoption of a large-scale high-resolution PS network in existing solutions frustrates the real-world deployment of hybrid beamforming architectures in mmWave small cell networks, where both base stations and mobile devices have strict limitations on the power consumption and hardware complexity.
Recently proposed partially-connected architecture can reduce the number of PSs and improve the energy efficiency \cite{Gao JSAC 16}. However, the cost of PSs is still unaffordable when a large-scale antenna array is adopted (even trading off with some performance loss).

Targeting at real-world deployment, this article introduces a novel hardware-efficient hybrid precoding/combining architecture. Inspired by recent works \cite{Rial Access}-\cite{Dai ICC 17}, we utilize a limited number of simple phase over-samplers (POSs) and a switch (SW) network to realize the analog beamformer, which can achieve maximum hardware efficiency as well as maintain satisfactory spectral efficiency performance.
The POS is implemented by a simple circuit and can simultaneously output several parallel signals with different phases. With the aid of a simple switch network, the analog precoder/combiner is realized by feeding the signals with appropriate phases to antenna arrays or RF chains.
After presenting the details of this novel POS-SW-based hybrid beamforming architecture, we analyze the design challenges and present potential solutions, especially the precoder/combiner design and the channel estimation mechanism.
Then, the advantages of the proposed architecture for hybrid beamforming in mmWave MIMO systems will be illustrated by simulation studies.
Finally, after highlighting the future research trends including potential performance improvements and technical extensions of the proposed scheme, some conclusion remarks are provided.

\section{Structure of Hardware-Efficient Hybrid Beamformer}
\label{sc:system model}


We consider a typical point-to-point mmWave massive MIMO system as an example, where the transmitter employs $N_t$ antennas and $N^{RF}_t$ RF chains  to simultaneously transmit $N_s$ data streams to the receiver which is equipped with $N_r$ antennas and $N^{RF}_r$ RF chains.

\subsection{Traditional PS-based Hybrid Beamforming Structure}

\begin{figure}[!t]
\centering
\includegraphics[width= 6.0 in]{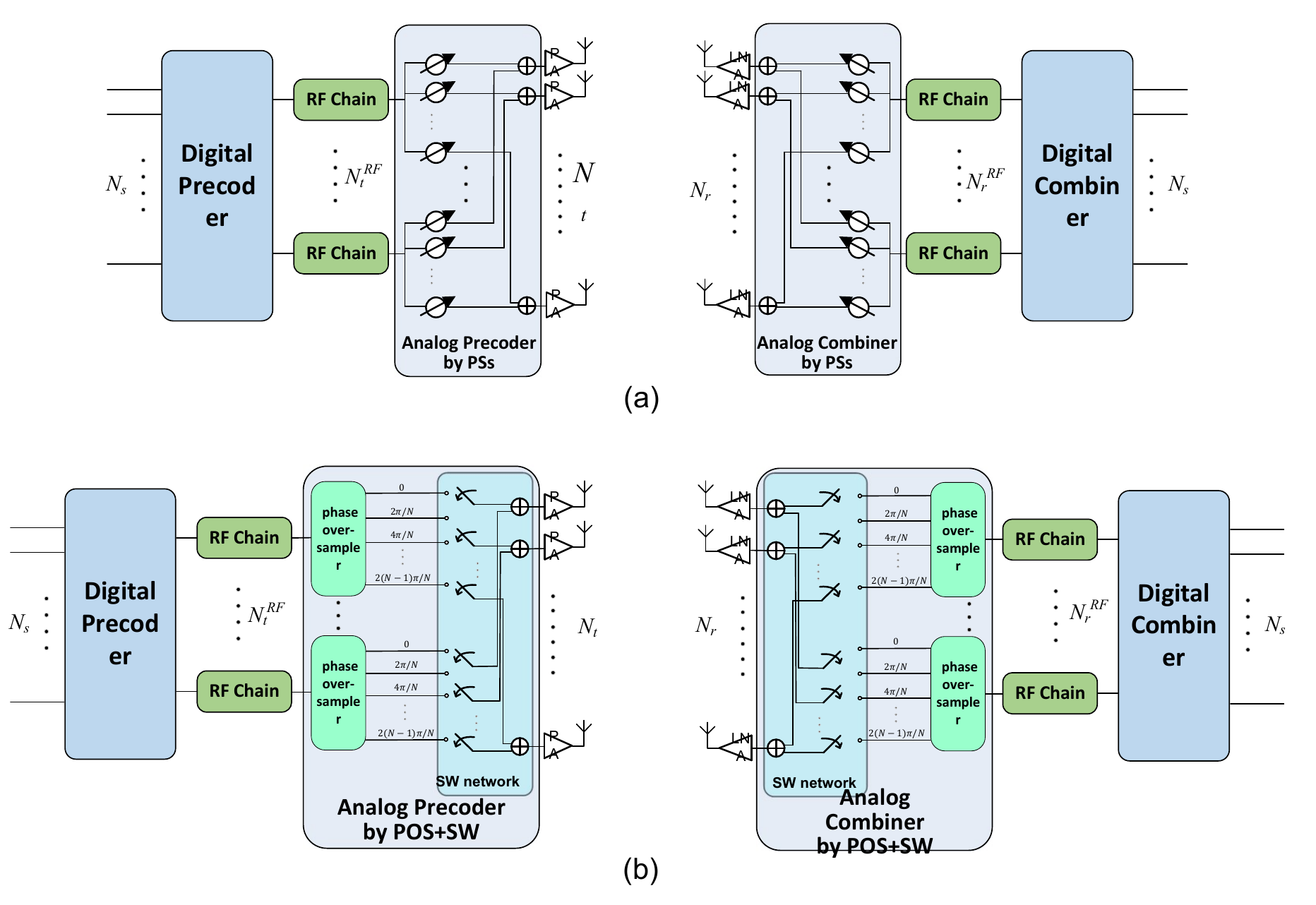}
\caption{(a) Traditional PS-based hybrid beamforming architecture. (b) Proposed POS-SW-based hybrid beamforming architecture.}\label{fig:system_model}\vspace{-0.0 cm}
\end{figure}

In the traditional hybrid precoding/combining architecture, as shown in Fig. \ref{fig:system_model}(a),
the transmitted symbols are first processed by a baseband digital precoder $\mathbf{F}_{BB} \in \mathbb{C}^{N_t^{RF}\times N_s}$, then up-converted to the RF domain via $N_t^{RF}$ RF chains before being precoded with an analog precoder $\mathbf{F}_{RF}$ of dimension $ N_t \times N_t^{RF}$. While the baseband digital precoder $\mathbf{F}_{BB}$ enables both amplitude and phase modifications, the elements of the analog precoder $\mathbf{F}_{RF}$, which are implemented by PSs, have a constant amplitude $\frac{1}{\sqrt{N_t}}$ and $N$-resolution quantized phases in practice. Obviously, a larger $N$ provides a finer resolution of PSs (i.e. a finer resolution of beamformer) and potentially better performance, but also results in higher hardware complexity and power consumption.
Similarly, the receiver employs an analog combiner $\mathbf{W}_{RF}$ under the same constraint as $\mathbf{F}_{RF}$ and a digital combiner $\mathbf{W}_{BB}$ to process the received signal. Let $\mathbf{H}$ denote the channel matrix, $\mathbf{s}$ represent the transmitted signal vector, $P$ represent transmit power, and $\mathbf{n}$ represent the received complex Gaussian noise vector. Then, the received signal can be expressed as
\begin{equation}
\mathbf{\widehat{s}} = \sqrt{P}\mathbf{W}^H_{BB}\mathbf{W}^H_{RF} \mathbf{H} \mathbf{F}_{RF}\mathbf{F}_{BB}\mathbf{s} + \mathbf{W}^H_{BB}\mathbf{W}^H_{RF}\mathbf{n}. \label{eq:received signal 2}
\end{equation}

This traditional PS-based hybrid beamforming structure requires a large number of PSs to implement analog beamforming.
For example in the fully-connected structure as shown in Fig. \ref{fig:system_model}(a), each RF chain is connected to all antennas via a large-scale PS network.
Obviously, this large-scale PS network will cause high power consumption and hardware implementation difficulty.
Moreover, the impractical assumption of high-resolution PSs in existing solutions frustrates the real-world deployment of hybrid beamforming architectures. The power consumption and cost of the PS at mmWave frequency band are proportional to its resolution. For example, a 4-bit (i.e. $N=16$) resolution PS requires 45-106 mW, while a 3-bit (i.e. $N=8$) resolution PS needs only 15 mW  \cite{Rial Access}.
Therefore, the hardware limitation, high power consumption and other problems in the traditional PS-based hybrid beamforming structures motivate us to seek for a hardware-efficient hybrid beamforming solution by reducing the number of PSs and the resolution of PSs while maintaining the performance of mmWave MIMO systems.

\subsection{Hardware-Efficient PS-SW-based Hybrid Beamforming Structure}

To overcome the shortcomings of the traditional hybrid beamforming and achieve a better trade-off between the hardware complexity and the system performance, we introduce a novel hardware-efficient structure in Fig. \ref{fig:system_model}(b). Taking the transmitter as an example, instead of using a large-scale PS network, each RF chain is connected to a phase over-sampler (POS) which can simultaneously output $N$ parallel signals with different phases as $0, \frac{2\pi}{N}, \ldots, \frac{2 (N-1) \pi}{N}$. For example, if $N=2$, the two output signals have binary phases of $0$ and $\pi$; if $N=4$, the four output signals have quaternary phases of $0$, $\frac{\pi}{2}$, $\pi$, and $\frac{3\pi}{2}$.
Then, the analog beamformer can be easily implemented by feeding the signal with appropriated phase to each antenna via a simple switch (SW) network.

\begin{figure}[!t]
\centering
\vspace{-0.0 cm}
\includegraphics[width= 6.0 in]{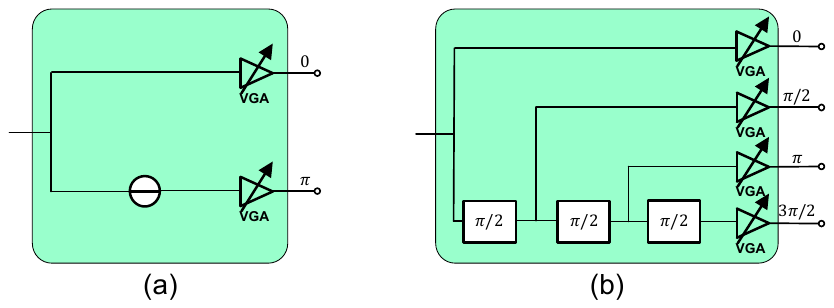}\vspace{-1.5 cm}
\caption{Examples of phase over-sampler at transmitter side. (a) Binary phase over-sampler; (b) Quaternary phase over-sampler.}\label{fig:PO}
\end{figure}

This POS-SW-based hybrid beamforming architecture has low hardware complexity and power consumption. First of all, the transmitter only needs $N_t^{RF}$ POSs, which have simpler hardware implementation than conventional digitally-controlled PSs.
For example, the two-phases (binary) POS can be easily realized by an inverter \cite{Dai ICC 17} as shown in Fig. \ref{fig:PO}(a); the four-phases (quaternary) POS can be implemented using a sequence of phase-shifting stages \cite{Poon 12} as shown in Fig. \ref{fig:PO}(b).
For the narrowband systems, simple microstrip delay-lines can be utilized to realize the phase-shifting stages.
This scheme has significant advantage of low hardware complexity, but suffers from the phase non-linearity problem (i.e. shifting different phases at different frequencies).
Therefore, for the wideband scenarios, a bank of constant (nun-tunable) LC-based wideband PSs should be employed to construct the phase-shifting stages to provide better performance in phase linearity \cite{Alkhateeb WCL 16}.
In addition, variable gain amplifier (VGA) is usually adopted to provide different gain compensation for different phases, which are connected with different number of antennas and may have different circuit loads.
Since only a limited number of POSs are employed, the overall hardware complexity is low and the power consumption is small.
Moreover, a switch circuit typically has $P_{SW} = 5$mW and the switch network has dramatically lower power consumption than the PS network.
Therefore, instead of using a large number of PSs, this POS-SW-based hybrid beamforming architecture can considerably reduce hardware complexity and power consumption.

\section{Design Challenges}
In this section, we discuss two main challenges in realizing the POS-SW-based hybrid beamforming  and introduce potential solutions which can tackle these challenges.

\subsection{Low-Complexity Precoder and Combiner Designs}

The objective of hybrid precoder and combiner design  is to maximize the spectral efficiency of this POS-SW architecture. The main difficulty comes from the low-resolution phase constraints of the analog beamformers.
The optimal exhaustive search algorithm has exponential complexity in the number of antennas and is definitely impractical for real-world implementation.
Thus, low-complexity designs of hybrid precoder and combiner for the POS-SW architecture to achieve satisfactory performance is a critical issue.

Recently, several hybrid beamforming design algorithms for low-resolution PS schemes have been proposed, which essentially have similar objective function as the POS-SW architecture.
In \cite{Chen 17}, the authors proposed cross-entropy minimization based analog beamformer design algorithms, where a large number of candidate beamformers are randomly generated and then iteratively refined to minimize the cross-entropy. Nevertheless, the performance will experience serious degradation when the number of selected candidates or iterations is not sufficiently large. An alternative codebook-free hybrid beamforming designs with discrete phases PSs were investigated in \cite{Sohrabi 16}. The authors first derived the optimal analog precoder with infinite-resolution phases, then directly quantized the phase term of each element to a finite set. Although this approach reduces the complexity of the hybrid beamforming designs, it cannot always maintain satisfactory performance  when the resolution of PSs is very low.
More recently, a novel joint hybrid precoder and combiner design algorithm was introduced in \cite{Wang JSTSP 17}, which has low complexity and near-optimal performance. For the design of hybrid beamformer with binary POSs ($N=2$), a small set of candidate codebooks are constructed based on rank-1 approximation, from which the optimal analog beamformers can be selected with polynomial complexity in the number of antennas.
For the case of $N>2$, a low-complexity phase matching algorithm was proposed to iteratively optimize the low-resolution phase term of each element of analog beamformer. It has been verified that fast convergence within three iterations can be guaranteed.

\subsection{Channel Estimation with POS-SW-based Hybrid Beamforming}

The hybrid precoder and combiner design requires full knowledge of channel state information (CSI), which is difficult to be obtained in mmWave MIMO systems since the channel is intertwined with analog beamformers and the baseband has no direct access to the entries of channel matrix.
Thanks to the specific sparse characteristic of mmWave channels in the angle domain, compressed sensing based approaches are often leveraged to implement efficient channel estimation by exploring the channel sparsity in mmWave system.
The seminal work \cite{Alkhateeb JSTSP 14} proposed an adaptive compressed sensing based channel estimation algorithm in conjunction with closed-loop beam training mechanism which needs a feedback link
between the transmitter and receiver.
In this type of approaches, hierarchical multi-resolution codebooks are designed to construct training beamformers with different beamwidths.
In order to accurately generate beams with different beamwidths, the phase resolution of the analog components should be sufficiently high.
However, those existing algorithms may be not applicable for the proposed POS-SW-based architectures due to the low resolution of POSs. Therefore,  hierarchical multi-resolution codebook design for POS-SW scheme with low-resolution phases needs to be studied to construct beams with adaptive beamwidths and sufficient gain within the coverage.

On the other hand, for the open-loop systems which do not have a feedback link, the mmWave channel estimation problem was formulated as sparse signal reconstruction problem \cite{Rial Access}, which can be solved via a widely used orthogonal matching pursuit (OMP) solution.
The training sequences of precoding/combining vectors can be designed using either pseudo-random sequences or deterministic measurement matrices.
It has been verified that the deterministic matrices enjoy lower mutual coherence of the training sequences and offer an advantage for channel estimation. Therefore, fast design of deterministic training beamforming matrices with discrete entries (e.g. $\{ \pm 1\}$ or $\{ \pm 1, \pm j\}$) should be studied for channel estimation of low-resolution POS-SW-based architecture.
In short, efficient channel estimation for low-resolution POS-SW-based scheme is highly critical for mmWave MIMO communications, permeating through the future communication technologies.

\section{Performance Illustrations}

\begin{figure}[!t]
\centering
  \includegraphics[width=4.5 in]{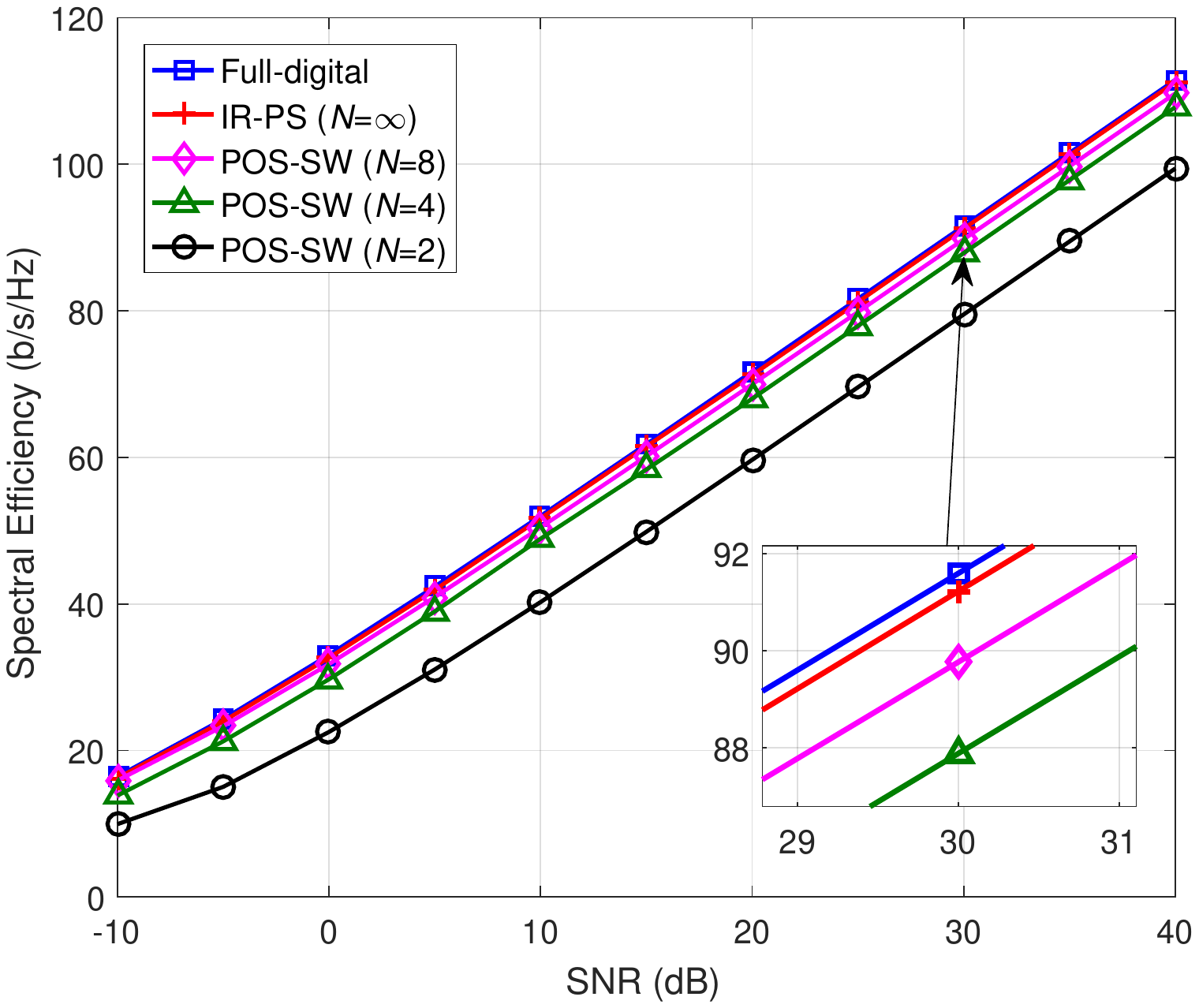}
  \vspace{-0.4 cm}
  \caption{Spectral efficiency versus SNR ($N_t = N_r = 64$, $N_t^{RF} = N_r^{RF} = N_s=6$).}\label{fig:SE_vs_SNR}
\end{figure}

In this section, we illustrate the performance of the POS-SW-based hybrid beamforming architecture and compare with the traditional PS-based hybrid and full-digital beamforming approaches. We adopt clustered mmWave channel model with $10$ scattering clusters, each of which contributes $5$ propagation paths \cite{Ayach TWC 14}.
The angles of arrival (AoA) and angles of departure (AoD) within a cluster are assumed to be Laplacian-distributed with angle spreads of $0.5^\circ$.
The mean AoD of a cluster is assumed to be uniformly distributed over $[0,2\pi]$, while the mean AoA of a cluster is uniformly distributed over an arbitrary $\frac{\pi}{3}$ sector.
We first assume that the transmitter and receiver are both equipped with $64$-antenna ULAs and the numbers of RF chains and data streams are $N_t^{RF}= N_r^{RF}=N_s=6$, respectively.
Fig. \ref{fig:SE_vs_SNR} illustrates the spectral efficiency performance of the POS-SW-based hybrid beamforming architecture as a function of signal-to-noise-ratio (SNR).
Recently proposed algorithm in \cite{Wang JSTSP 17} is adopted for designing the precoders and combiners with phase resolutions $N=2$, $N=4$, and $N=8$, respectively.
For comparison purposes, we also include the spectral efficiency performance of full-digital scheme with optimal SVD-based beamforming design and infinite-resolution-PS-based (IR-PS) hybrid scheme (i.e. $N= \infty $) with phase extraction (PE-AltMin) beamforming design algorithm \cite{Yu JSAC 16}.
We can observe that the POS-SW-based hybrid beamforming with $N=4$ phase resolution can achieve satisfactory spectral efficiency performance close to the full-digital beamforming scheme and the hybrid beamforming scheme with infinite-resolution PSs. Finer phase resolution (e.g. $N=8$) can improve the spectral efficiency performance but may suffer from higher hardware complexity and cost.
For the case $N=2$ in which the hybrid beamforming architecture has lowest hardware complexity, there is notable performance loss due to the extreme low resolution of POSs.

%
%
%

\begin{figure}[!t]
\centering
  \includegraphics[width= 4.5 in]{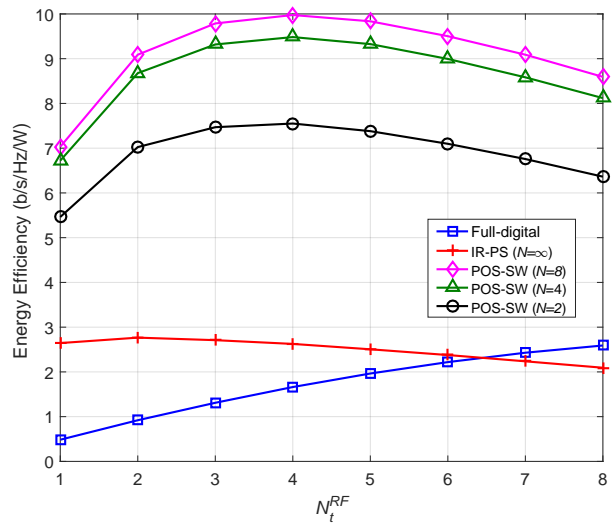}
  \vspace{-0.4 cm}
  \caption{Energy efficiency versus number of RF chains $N_t^{RF}$ ($N_t=N_r=64$, $N_s = N_r^{RF}=N_t^{RF}$).}\label{fig:EE_vs_Ns}
\end{figure}

In Fig. \ref{fig:EE_vs_Ns}, we turn to illustrate the energy efficiency of different beamforming architectures. The energy efficiency is defined as the ratio of spectral efficiency to the total power consumed at the transmitter side. The total power consists of the power  for realizing beamformer and the power for transmitting signals. Note that for different architectures, the power for beamforming is consumed by different components. For the full-digital scheme, the power consumption for beamforming comes from the baseband processor and $N_t$ RF chains.
For the IR-PS-based hybrid beamforming architecture, the beamforming power consists of the consumptions by the baseband processor, $N_t^{RF}$ RF chains, and $N_tN_t^{RF}$ PSs.
For the proposed POS-SW-based hybrid beamforming architecture, the beamforming power is consumed by the baseband processor, $N_t^{RF}$ RF chains, and $N_tN_t^{RF}$ SWs. We should emphasize that in the POS-SW structure, a small number of POSs are implemented by simple  inverters or microstrip delay-lines as shown in Fig. \ref{fig:PO} and have negligible power consumption. Therefore, the power consumption of POS-SW-based  architecture is the same as that of IR-PS-based structure except for replacing the consumption of the PS network by the SW network.
In the simulation, the power consumptions of the baseband processor and each RF chain are set as $P_{BB}=200$mW and $P_{RF}=300$mW \cite{Dai ICC 17}, respectively.
Moreover, the power consumption of each PS is $P_{PS}=40$mW and the power consumption of each SW is $P_{SW}=5$mW \cite{Dai ICC 17}. Finally, we also assume the transmit power is $P=500$mW, which is a typical setting in existing literatures. Fig. \ref{fig:EE_vs_Ns} illustrates the energy efficiency performances versus the number of RF chains. It can be observed that the proposed POS-SW architecture enjoys significant energy efficiency advantages compared with the conventional full-digital and PS-based hybrid schemes.

%

\begin{figure}[!t]
  \includegraphics[width= 6.5 in]{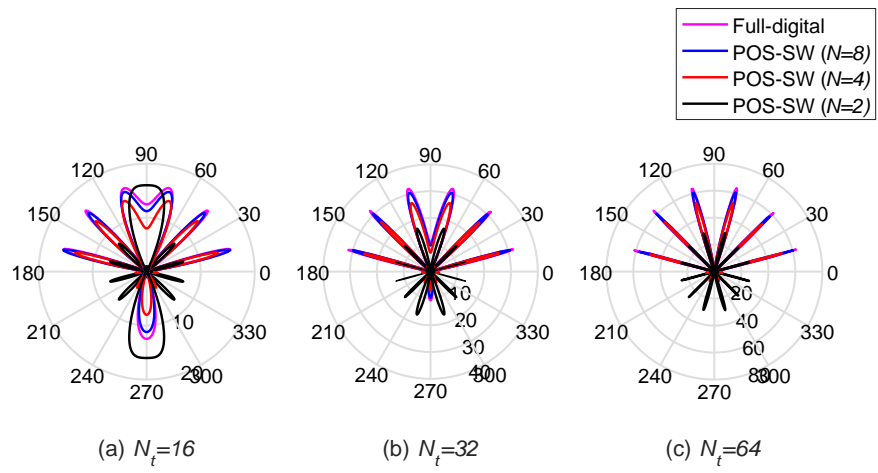}
  \vspace{-1.4 cm}
  \caption{Beam patterns with $15^{\circ}$, $45^{\circ}$, and $75^{\circ}$ DoDs.}\label{fig:pattern}
\end{figure}

Fig. \ref{fig:pattern} shows examples of beam patterns of the proposed scheme with different phase resolution, directions of departure (DoDs), and number of antennas. From this figure we find that the proposed scheme with $N=4$ and $N=8$ phase resolution can generate good beam patterns.
For $N=2$ case, however, the beam pattern has small main-lobe and difficulty to steer at $75^{\circ}$ with small number of antennas ($N_t=16$) due to the extremely low phase resolution.
This observation also verifies the results of the spectral efficiency performance evaluation in Fig. \ref{fig:SE_vs_SNR}.

\section{Future Research Trends}

\subsection{Intelligent Connectivity Mechanisms for SW Networks}

\begin{figure}[!t]
\centering
\includegraphics[width=6.0 in]{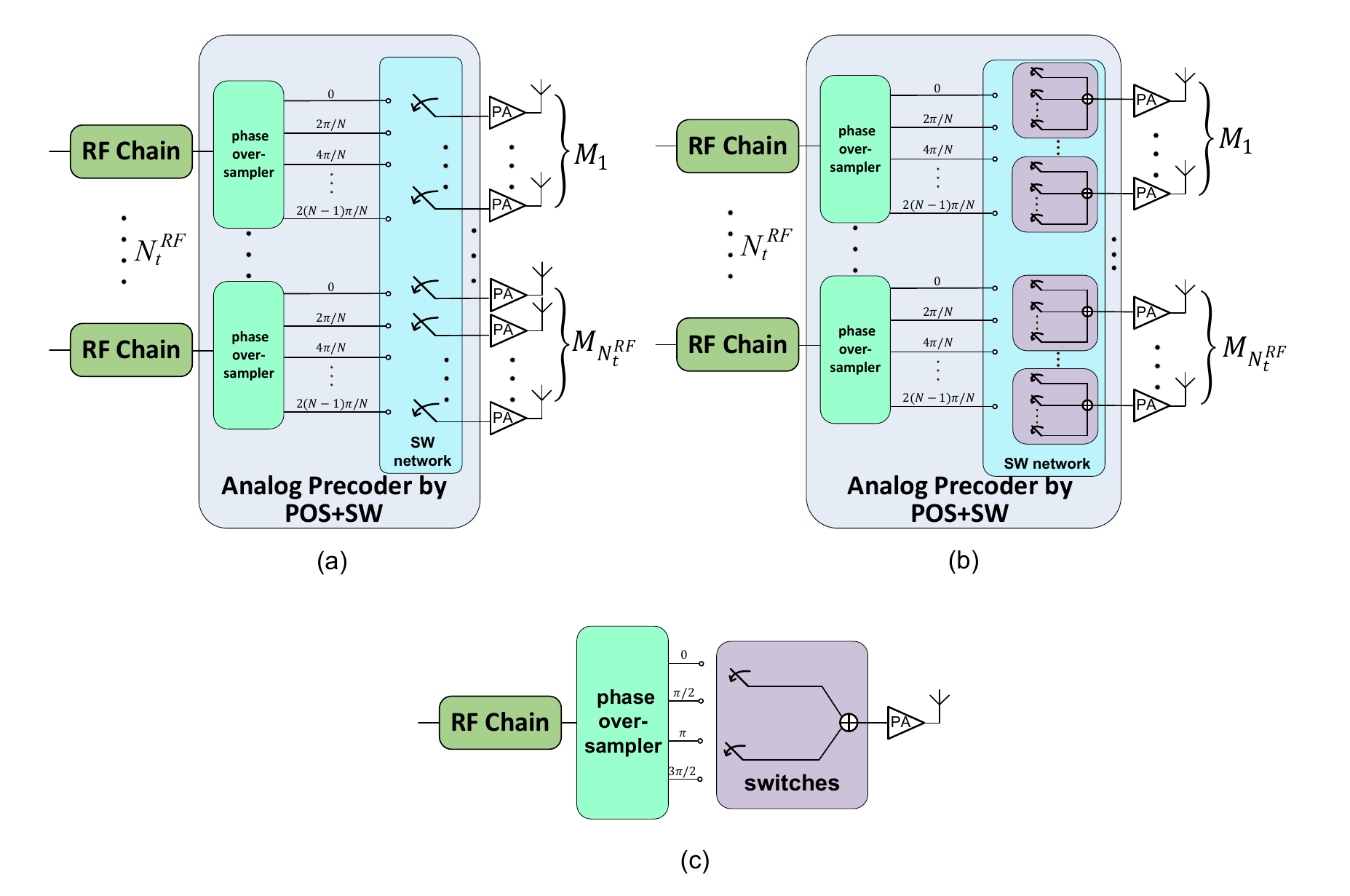} \vspace{-1.4 cm}
\caption{(a) Partially-connected structure with $N_t$ switches. (b) Partially-connected structure with more than $N_t$ switches. (c) An example of dynamic connection structure with a quaternary POS.}\label{fig:dynamic system}
\end{figure}

The POS-SW-based hybrid precoding/combining structure can provide more degrees of freedom (DoF) in the analog domain by intelligently connecting the SW network.
Fig. 1(b) introduces a representative fully-connected structure in which the phase-controlled signals of each POS are fed to all antennas to achieve full array gain.
Certainly, it is of great interest to utilize the flexibility of the SW network to explore more efficient structural connections.
Partially-connected structure as illustrated in Fig. \ref{fig:dynamic system}(a) is another typical hybrid beamforming scheme, in which the signals of each POS are routed to a subset of antennas instead of all antennas.
In conventional partially-connected structures, antenna array is equally divided into $N_t^{RF}$ non-overlapping sub-arrays.
In order to achieve better spectral efficiency and energy efficiency performances, it would be more efficient to allow different sub-array sizes as well as overlapped sub-arrays.
This motivates us to seek for intelligent connection approaches that can dynamically select the antenna sub-arrays and connect the SW network based on the characteristics of mmWave MIMO channels.

Moreover, with the flexibility of combining more than one phased-controlled signal of each POS as shown in Fig. \ref{fig:dynamic system}(b), we are able to adjust the signals with more phases and magnitude levels and provide more DoF to potentially achieve better performance.
Considering an example in Fig. \ref{fig:dynamic system}(c) with a $N=4$ POS and two switches, the output signals may have eight different phases (e.g., $0$, $\frac{\pi}{4}$, $\frac{\pi}{2}$, $\frac{3\pi}{4}$, $\pi$, etc.) and three different magnitudes. While the POS can only offer signals with constant magnitude and 4 different phases, this solution can provide more DoF on adjusting both phases and magnitudes to obtain more accurate analog beam.
This intelligent connectivity mechanism can be applied to both fully-connected and partially-connected architectures to improve the spectral efficiency performance.
However, the potential performance improvement is achieved at the expense of a more complicated switch network and more power consumption.
Therefore, investigating the intelligent connection architecture and seeking efficient algorithms for channel estimation and beamforming design are important ongoing research challenges to balance the trade-off between the spectral efficiency performance and hardware structural constraints.

\subsection{Extension to Multi-user mmWave Systems}
Based on the foundation of the point-to-point hybrid precoding/combining architecture, the extension to multi-user mmWave systems is another important research topic. In the multi-user hybrid precoding/combining systems, each user should be assigned an analog beamformer to achieve maximum channel gain as well as mitigate inter-user interference. The baseband digital processing will further suppress the interference between users.
However, the POS-SW-based hybrid beamforming structures may suffer from the difficulty of finely steering the analog beams, which will lead to stronger inter-user interference.
Therefore, it is of great interest to seek efficient beamforming designs to reduce the interference and improve the efficiency of a multi-user mmWave system.
Moreover, efficient channel estimation mechanisms and algorithms for the multi-user mmWave systems are also desired.

\subsection{Combining with Low-Precision ADCs and DACs}

In mmWave systems, the sampling rate of the ADCs and DACs will dramatically scale up due to orders-of-magnitude wider bandwidth (e.g., $> 1$GHz).
Unfortunately, high-speed (e.g., $>2$GS/s) high-precision (e.g., 8-12 bits) ADCs/DACs are very power-hungry and costly.
The hybrid beamforming architecture can significantly reduce the cost and power requirement by employing a limited number of RF components and ADCs. An alternative solution for mmWave systems is making use of low-precision ADCs, especially one-bit ADCs/DACs which have extremely low hardware complexity and power consumption (e.g., the power consumption of a one-bit 3.6GS/s ADC is only $15mW$).
While the low-precision ADCs/DACs are advantageous in terms of power consumption and hardware complexity, they
generally cause severe performance loss due to larger quantization error.
For the transmission over signal-antenna additive white Gaussian noise (AWGN) channel, there is 0.22dB spectral efficiency loss at low SNR (0dB) when 2-bit ADCs are employed and 0.7dB loss at high SNR (20dB) with 3-bit ADCs.
Using larger antenna arrays in conventional MIMO systems can compensate for the spectral efficiency performance loss.
Clearly, it is worths investigating the combination of low-precision ADCs/DACs and low-resolution POS-SW-base hybrid beamformer.
The beamforming design and performance analysis of this integrated structure need to be studied, especially the closed-form expression of performance which can help us to obtain insights of important design issues, such as how to determine the appropriate quantities and resolution of ADCs/DACs and POSs in order to achieve an appropriate trade-off between hardware efficiency and system performance.

\subsection{Beamforming Design for Wideband mmWave Systems}
Most existing hybrid precoding/combining solutions are based on narrowband mmWave channels. However, we should emphasize that the mmWave systems are expected to operate in a wideband mode to enjoy the orders-of-magnitude wider bandwidth. The properties of wideband mmWave channels, such as frequency selectivity, path loss, delay spread, angel spread and the number of clusters, may be quite different from the narrowband mmWave channels. Therefore, the existing hybrid beamforming designs for the narrowband mmWave channels are not suitable for the wideband mmWave systems.
Moreover, when the simple delay-lines are utilized in
the POS to generate signals with different phases, we should be aware of the non-linearity
problem in which signals with different frequencies will have different phase shift through a
delay-line. Even for LC-based PSs, the good phase linearity is also a critical performance parameter which should be concerned.
Besides, when a large-scale antenna array is employed, the wideband signal will be sensitive to the physical propagation delay across the large array aperture. This so-called ``spatial-wideband effect'' is another important issue which is usually ignored in the past.
Therefore, the research on the hybrid beamforming design, analysis on performance loss due to non-linearity problem and spatial-wideband effect, and pre-compensation strategy should be conducted in the future works for the expansion of POS-SW-based hybrid beamforming structures in the wideband mmWave systems.

\section{Conclusions}
\label{sc:Conclusions}
Hybrid precoding and combining will be an important component of future mmWave MIMO communication systems.
This article is dedicated to introducing a novel hardware-efficient hybrid beamforming structure which utilizes a limited number of simple POSs and a switch network to implement analog beamforming.
We analyzed design challenges and presented potential solutions of precoder/combiner design and channel estimation for this hybrid beamforming architecture.
Future research directions were also discussed.
It is expected that this hardware-efficient hybrid beamforming structure will play an important role in future mmWave MIMO communication systems.

\begin{IEEEbiographynophoto}
{Ming Li} (S'05, M'11, SM'17) received the M.S. and Ph.D. degrees in electrical engineering from State University of New York at Buffalo in 2005 and 2010, respectively. He is presently an Associate Professor in Dalian University of Technology, China. His current research interests are in the general areas of communication theory and signal processing with applications to mmWave communications, secure wireless communications, cognitive radios and networks, data hiding and steganography.
\end{IEEEbiographynophoto}

\begin{IEEEbiographynophoto}
{Zihuan Wang}(S'17) received the B.S. degree in communication engineering from Dalian University of Technology, Dalian, China, in 2017. She is now studying toward the M.S. degree with the School of Information and Communication Engineering, University of Technology, Dalian, China.
Her current research interests are focused on signal processing in mmWave communications, MIMO communications, and multicasting systems. She received the National Scholarship in 2017.
\end{IEEEbiographynophoto}

\begin{IEEEbiographynophoto}
{Hongyu Li} is currently studying toward the B.S. degree in electrical and information engineering from Dalian University of Technology, Dalian, China. Her current research interests include signal processing in mmWave communications, massive MIMO, and antenna selection. She received the National Scholarship in 2016 and 2017, respectively.
\end{IEEEbiographynophoto}

\begin{IEEEbiographynophoto}{Qian Liu} (S'09, M'14) is an Associate Professor at the Dept. of Computer Science and Technology, Dalian University of Technology, China. Her current research interests include smart vehicular communications, immersive multimedia processing and communications. She provides services to the IEEE Haptic Codec Task Group as a secretary for standardizing haptic codecs in the Tactile Internet. She also served as the Technical Committee Co-Chair of HAVE 2017 and HAVE2018.
\end{IEEEbiographynophoto}

\begin{IEEEbiographynophoto}{Liang Zhou} is a professor in Nanjing University of Posts and Telecommunications, China.
His research interests are in the area of multimedia communications and networks. He currently serves as an editor for IEEE Transactions on Circuits and Systems for Video Technology, IEEE Transactions on Multimedia, and IEEE Network. He also serves as Co-Chair and Technical Program Committee (TPC) member for a number of international conferences and workshops.
\end{IEEEbiographynophoto}

\clearpage

\clearpage

\clearpage

\clearpage

\clearpage

\clearpage

\end{document}